# Towards Indonesian Speech-Emotion Automatic Recognition (I-SpEAR)

Novita Belinda Wunarso[*], Yustinus Eko Soelistio  
Information System Department  
Universitas Multimedia Nusantara  
Serpong, Indonesia 15811  
Email: {novita.belinda[*], yustinus.eko}@(student.[*])umn.ac.id

*Abstract*—Even though speech-emotion recognition (SER) has been receiving much attention as research topic, there are still some disputes about which vocal features can identify certain emotion. Emotion expression is also known to be differed according to the cultural backgrounds that make it important to study SER specific to the culture where the language belongs to. Furthermore, only a few studies addresses the SER in Indonesian which what this study attempts to explore. In this study, we extract simple features from 3420 voice data gathered from 38 participants. The features are compared by means of linear mixed effect model which shows that people who are in emotional and non-emotional state can be differentiated by their speech duration. Using SVM and speech duration as input feature, we achieve 76.84% average accuracy in classifying emotional and non-emotional speech.

*Keywords—speech-emotion recognition, Daubechies wavelet, Linear mixed effect, SVM.*

## I. Introduction

Technological development enables human to interact with fellow human and electronic devices such as virtual assistant on smart phones or computers. This technology enables computers to understand users' emotion and apply appropriate feedback in a manner that is much easier for users to accept. Emotion recognition itself is an essential component of affective computing [1] and the media through which human emotion can be recognized includes speech signals, facial images, gestures, bio-signals and skin temperature [2]. Many efforts have been made to recognize various emotions from speech using different extracted features and classifiers [3-5]. However, there are still some disputes on which vocal features can identify certain emotions, and also most of the research occupies different languages and cultural backgrounds [3-5]. Emotion is very much dependent on culture, environment, and the pre-emotional state of human [6] which make the result of existing studies could be applied universally across cultures. In our knowledge, there are only a few studies in this field that covers Indonesian language. Hence, further research using Indonesian language is necessary. In the following, we discussed three recent methods on speech-emotion recognition (SER) and how they relate to our study.

A method by Pan et. al. [3] attempt to recognize three emotional states: happy, sad, and neutral was done by combining various features such as energy, pitch, linear predictive spectrum coding (LPCC), mel-frequency spectrum coefficients (MFCC), and mel-energy spectrum dynamic coefficients (MEDC). They use the Berlin Database of Emotional Speech and self-built Chinese emotional database to train the support vector machine (SVM) classifier. They found that different combination of emotional characteristic features could obtain different recognition rate and these features are sensitive to languages being used.

Method by Hamidi and Mansoorizade [4] attempt to classify six emotions (anger, disgust, fear, sadness, and happiness) from Persian Emotional Speech Database. This database was built from actor or actress speech from more than 60 different films. Four features (energy, pitch, MFCC coefficients, its first derivative, and the minimum and maximum range of data) are extracted to classify the data using neural network. It can reach accuracy of 78% on classifying the data.

Another speech emotion recognition was done on Marathi speech database using MFCC and wavelet coefficients from discrete wavelet transform (DWT) as features [5]. The SVM classifier was trained to recognize anger, boredom, fear, happy, sad, and neutral emotion. The author claims that the performance of system for Marathi speech database was promising.

The three aforementioned methods successfully classify emotions using spectrum extraction techniques and non-linear classifier. Our study uses similar spectrum extraction (DWT, [5]) and both neural network and SVM [3, 5] to classify three emotions (happy, sad, and neutral) in Indonesian language database. We use simple three features: amplitude, duration or length of speech signal, and approximation coefficients derived from DWT. We employ linear mixed effect models to analyze the effect of the emotion on these features and also the effect of other factors such as subject, gender, and word.

The rest of this paper is organized as follows. Section II explains our data collection method, which is followed by the research methodology in Section III. The implementation and results are presented in Section IV, the discussions and conclusions are given in Section V and VI respectively.

## II. DATA COLLECTION

Other studies mostly use acted emotional speech databases. [4] made the Persian emotional speech database from actor or actress speech from more than 60 different films. It contained 2400 pieces of emotional speech from both men and woman with different ages and included anger, disgust, fear, sadness, happiness and normal emotion. Other study [5] built the database from Marathi continuous sentences uttered in 6 emotions i.e. happiness, anger, neutral, fear, sadness, boredom by 10 speakers (5 female and 5 male). A set of 10 sentences uttered three times each by the speakers for all 6 emotions. Files of one speaker were presented to 10 naive listeners and only those sentences which emotions were identified by at least 80% of the listeners were selected.

In this research, we use elicited emotional speech to build the Indonesian speech-emotion database (I-SpeED). 40 participants with average age of 21.4 years old ($\sigma = 0.9$) consisting of 20 males and 20 females are randomly selected from Universitas Multimedia Nusantara. All participants are students and given souvenir and credit as part of coursework. The main requirement for participant is he/she does not speak in accent. Before starting the data collection procedure for each emotion, the participant is asked to fill a DES questionnaire [11] about his/her emotional state at that time (cf. [8]). The questionnaire consists of 16 emotions: surprise, anger, anxiety, calm, confusion, contempt, disgust, embarrassment, enthusiasm, fear, shame, happiness, interest, love, pride, and sadness to be filled with scales ranging from 1 (the participant does not feel the emotion at all) to 5 (the participant feels the emotion intensely) to monitor participants emotional states. Emotion elicitation is done by asking the participant to watch a film clip and to share his/her personal experiences regarding the targeted emotion. Film clips are used as they allow unrestrained body movements which might be used to elicit strong responses (even negative ones) without creating a sense of harm or ethical violation [8] on recalling and telling past emotional experiences [9-10]. After watching the film clip and sharing his/her emotional experience, the participant is asked to fill the questionnaire once again and begin the I-SpeED recording.

TABLE I.   WORDS IN THE I-SPEED. THE WORDS IN THE BRACKET IS THE LOOSE TRANSLATION IN ENGLISH.

| Words | | |
|---|---|---|
| Perut (*stomach*) | Kamera (*camera*) | Tangan (*hand*) |
| Mobil (*car*) | Darat (*land*) | Momen (*moment*) |
| Permen (*candy*) | Waktu (*time*) | Batik (*batik*) |
| Hujan (*rain*) | Rumah (*house*) | Negara (*country*) |
| Akris (*actrees*) | Ikan (*fish*) | Baja (*steel*) |
| Kelapa (*coconut*) | Album (*album*) | Surat (*letter*) |
| Kabin (*cabin*) | Pasar (*market*) | Calon (*candidate*) |
| Garam (*salt*) | Motor (*motorcycle*) | Payung (*umbrella*) |
| Soda (*soda*) | Swasta (*private*) | Warga (*citizen*) |
| Eksport (*export*) | Acara (*event*) | Daerah (*region*) |

To elicit happiness, neutral, and sadness emotion, we use short clips from three films: Wall-E, Searching for Bobby Fischer, and Saving Private Ryan by following the instruction from [11]. We adapt the methods in building the SoNaR database of Dutch words [12] by taking 200 nouns from 21 news articles gathered from various news sites such as Detik, Okezone, and Liputan6 as candidates of neutral word. The neutrality of these words is evaluated by five people, who don't participate as participants, to be evaluated. Furthermore, 30 words are chosen randomly from the list of words which are stated as neutral by at least three of the evaluators (none of these evaluators are chosen as participants) to build the I-SpeED (Table I). The I-SpeED is publicly available for research purpose, and can be obtained by contacting one of the authors.

To validate whether the emotions felt by the participants are synchronous with the targeted emotions, all participants DES scores after the stimulation (i.e. watching film clip) and prior the experiment are compared by means of Anova. We conclude the data is valid only if: (1) the DES score between post and prior the experiment are significantly different, (2) the post stimulation DES score of the targeted emotion are significantly different than the others of non-targeted emotions, and (3) the utterances are clear. We find that the data from 38 participants are valid (all needed $p \leq 0.05$) which consist of 3420 of voices data.

We use Praat [13] to record voice data with 16KHz/16bit sampling rate. The microphone is placed around 25-30 centimeters from the participant's face (cf. [5]).

## III. METHODOLOGY

The emotions are classified by using features which show significant differences as the effect of emotion in linear mixed effect models. These features are given as input to neural network and SVM to classify speech data into happy, neutral, and sad emotion class. This section describes: (1) the feature extraction, (2) feature analysis, and (3) emotion classification.

### A. Feature Extraction

We extract three features from the voice data: the average of amplitude, average of approximation coefficients derived from one level decomposition of DWT using Daubechies wavelet db1 to db4, and the length of speech signal (the beginning until the end of the utterance). Each of these features shows the voice volume, frequency, and duration of speech respectively.

### B. Feature analysis

Linear mixed effect models are built to analyze the effect of emotion and gender as additional effect on extracted vocal features. The fixed effect of the model is emotion, while the random effect is gender. We use the lme4 library [14] in R to build the models and compare the full model (model with complete fixed and random effects) with the null models (model without fixed effect or the random effect) using the Anova. Any significant differences between the full and null models indicate the significant contribution of the respective effect on the model.

*C. Emotion Classification*

Neural network and SVM classifiers are employed to classify emotion from the speech data. To evaluate the performance of the classifiers, we implement 10-fold cross validation and calculate the average of the accuracy rate from all folds as the average performance of the classifiers.

## IV. Implementation and Results

*A. Features Selection*

By comparing the full model with the null model, the effect of emotion on vocal features is examined. We find no significant differences by adding emotion as the fixed effect of models with amplitude and approximation coefficients as response variables ($p > 0.05$) except on the duration as the response variable (Table II), hence we conclude that only utterance duration can be used for classification.

TABLE II. Linear Mixed Model of the full and null model comparison results

| Response variable | *p*-value |
|---|---|
| Amplitude | 0.09 |
| Duration | 6.214e-09 |
| Db1 | 0.09 |
| Db2 | 0.08 |
| Db3 | 0.07 |
| Db4 | 0.08 |

We also examine the effect of gender on vocal features, and find significant difference on the model with the duration as the response variable ($p = 4.4e - 07$).

Comparing the full model's intercept ($\beta$) adjustment points of the speech's duration on the three emotions, we find that the neutral and both sad and happy emotions show obvious difference ($\beta_{happy} = 9331.8 \pm 1420.2$, $\beta_{(neutral,happy)} = -282 \pm 56.5$, $\beta_{(sad,happy)} = 32.5 \pm 55.7$), which indicate the duration of neutral speech is shorter than the other two.

*B. Emotion Classification*

From the result of linear mixed effect models, it is known that the duration of speech signal can be used to differentiate between neutral speech and both sad and happy speech (we will call this as emotional and non-emotional speech for the rest of the article). We use this feature as the input of neural network and SVM to classify between the two types of speech. The neural network uses a single sigmoid neuron while the SVM uses polynomial kernel. The total number of data for emotional speech is twice (2280 data) than non-emotional speech (1140) as it contains combined data from the sad and happy speech.

The performance of the neural network and SVM are evaluated using method discussed in section III.C. The evaluation shows that the neural network and SVM perform 66.7% and 76.8% respectively (Table III). Although the neural network reach the accuracy of 66.7%, it is clear that this result is artificial since all samples are falsely classified as emotional speech. On the other hand, the SVM shows a more promising result where the precision and recall rate of the non-emotional speech is higher (60% and 67.1% respectively). Since, in our knowledge, this is the first exploration on classifying emotion based on Indonesian speech, then this research can be used as the first baseline for future SER researches based on Indonesian language.

TABLE III. Classification results using duration feature

| Output | Neural Network | | SVM | |
|---|---|---|---|---|
| **Input** | *Non-emotional* | *Emotional* | *Non-emotional* | *Emotional* |
| *Non-emotional* | 0 | 1140 | 684 | 456 |
| *Emotional* | 0 | 2280 | 336 | 1944 |
| Precision | 0% | 100% | 60% | 85.3% |
| Recall | - | 66.7% | 67.1% | 81% |
| Accuracy | 66.7% | | 76.8% | |

## V. Discussion

Although we do not to find any useful features for recognizing the three emotions, we find that the speech duration can be used to effectively differentiate the emotional speech from non-emotional speech. Using the speech duration, the SVM shows higher accuracy than neural network by correctly classifying 60% of the non-emotional and 85% of the emotional data. Neural network performs poorly with one dimensional input feature as there is no correctly classified non-emotional data. The poor classification of the non-emotional speech can be caused by the imbalance number of samples which makes the classifier more favorable to the larger data class (emotional class).

Furthermore we find that male significantly speaks faster than female ($p = 4.4e - 07$, $\mu_{male} = 0.49$ second, $\mu_{female} = 0.67$ second). This suggests a clue that duration based SER should be grouped based on the gender prior the recognition.

## VI. Conclusion

In this study we extract the amplitude, speech duration, and approximation coefficients of db1 to db4 from I-SpeED. Using linear mixed effect models, we find that there is a significant difference in the duration between the emotional (either happy or sad) and the non-emotional (neutral) speech. When used for classification using SVM, the evaluation shows that it achieves the average accuracy of 76.84%. We also find that men speak faster than women which make an interesting clue to classify SER based on gender for future studies.

## Acknowledgment

This study is supported by Universitas Multimedia Nusantara under grant number 078/LPPM-UMN/III/2016. The result of this study is part of the thesis by Novita Belinda Wunarso.